# A dispersion-engineered multi-pass cell for single-stage post compression of an Ytterbium laser

Laura Silletti,[1,*] Ammar bin Wahid,[1] Esmerando Escoto,[2] Prannay Balla,[2,3,4] Supriya Rajhans,[2,5] Katinka Horn,[6,7] Lutz Winkelmann,[2] Vincent Wanie,[1] Andrea Trabattoni,[1,8] Christoph M. Heyl,[2,3,4] and Francesca Calegari[1,9,10]

[1]Center for Free-Electron Laser Science CFEL, Deutsches Elektronen-Synchrotron DESY, Notkestr. 85, 22607 Hamburg, Germany
[2]Deutsches Elektronen-Synchrotron DESY, Notkestr. 85, 22607 Hamburg, Germany,
[3]Helmholtz-Institut Jena, Fröbelstieg 3, 07743 Jena, Germany
[4]GSI Helmholtzzentrum für Schwerionenforschung GmbH, Planckstraße 1, 64291 Darmstadt, Germany
[5]Friedrich-Schiller-Universität Jena, Max-Wien-Platz 1, 07743 Jena, Germany
[6]Department of Chemistry and Applied Biosciences, Laboratory of Physical Chemistry, ETH Zürich, Vladimir-Prelog-Weg 2, 8093 Zürich, Switzerland
[7]Center for Molecular and Water Science CMWS, Deutsches Elektronen-Synchrotron DESY, Notkestr. 85, 22607 Hamburg, Germany
[8]Institute of Quantum Optics, Leibniz University Hannover, Welfengarten 1, 30167 Hannover, Germany for Molecular and Water Science CMWS, Deutsches Elektronen-Synchrotron DESY, Notkestr. 85, 22607 Hamburg, Germany
[9]The Hamburg Centre for Ultrafast Imaging, Universität Hamburg, Luruper Chaussee 149, 22761 Hamburg, Germany
[10]Institut für Experimentalphysik, Universität Hamburg, Luruper Chaussee 149, 22761 Hamburg, Germany
*Corresponding author: laura.silletti@desy.de

Post-compression methods for ultrafast laser pulses typically face challenging limitations including saturation effects and temporal pulse break-up when large compression factors and broad bandwidths are targeted. To overcome these limitations, we exploit direct dispersion control in a gas-filled multi-pass cell, enabling for the first time single-stage post-compression of 150 fs pulses and up to 250 µJ pulse energy from an Ytterbium (Yb) fiber laser down to sub-20 fs. Dispersion-engineered dielectric cavity mirrors are used to achieve nonlinear spectral broadening dominated by self-phase-modulation over large compression factors and bandwidths at 98% throughput. Our method opens a route towards single-stage post-compression of Yb lasers into the few-cycle regime. © 2020 Optica Publishing Group

Ytterbium (Yb)-based laser systems are playing an increasingly important role in the field of ultrafast science. In contrast to Ti:Sapphire (Ti:Sa) laser systems, Yb-based lasers are power-scalable into the kilowatt (kW) regime while operating at high repetition rates [1], making them a valuable tool not only for average power-demanding applications, but also for improving the signal-to-noise ratio of laser-based experiments. However, amplifier gain bandwidth as well as gain-narrowing effects typically limit their pulse duration to >100 fs. A route to overcome this limitation is the use of spectral broadening techniques with subsequent post-compression, allowing for the generation of few-cycle pulses with energies reaching far into the mJ regime (mJ) [2-6]. In this context, gas-filled or solid-state-based multi-pass cells (MPC) have emerged as promising alternative to the more conventional hollow-core-fiber technique as they support high efficiency, great energy scaling options, low beam pointing susceptibility and compact setups [6,7]. Using a single-stage MPC, post-compression of Yb-lasers to about 30 fs has been already demonstrated with an overall transmission exceeding 96% [4,8]. Recently, various attempts have been made to enable direct compression of Yb-lasers into the few-cycle regime [9,10], for instance a double-stage approach has been used to reach compression of 1.2 ps pulses down to 13 fs. This result has not only been achieved in a cascaded arrangement of long MPC setups, but also with the use of metallic cavity mirrors to support sufficient spectral bandwidth, causing significant losses. Other recent approaches use cascaded post-compression schemes employing multiple stages [11] as well as fiber-based modal mixing schemes [12].

An important asset to mitigate general limitations of spectral broadening such as peak power degradation, self-steepening or temporal pulse break-up is the direct control of dispersion and nonlinearity, determining dispersion length $L_D$ and nonlinear length $L_N$, as defined e.g. in [13] and [14], respectively. The ratio between these two quantities $L_D/L_N$ determines the broadening regime. For $L_D/L_N \gg 1$, the pulse spreads quickly in time and the broadening process saturates. In contrast, for $L_D/L_N \ll 1$, i.e. approaching a dispersion-free or dispersion-balanced regime, self-phase modulation dominates and new frequencies are generated very efficiently without a severe impact on the temporal pulse shape. Dispersion-control has been exploited e.g. in bandgap hollow-core photonic crystal fibers (HC-PCFs), by means of gas pressure gradients [15,16] or using Kagomé-type structured fibers. In fact, spectra supporting sub-10 fs pulses have been demonstrated, however, at low pulse energy and at the cost of low energy transmission [17,18]. In addition, in particular when targeting large compression factors and broad bandwidth, dispersion-balanced regimes can hardly be reached with fiber-based spectral broadening methods. Here, MPCs can offer a solution, supporting dispersion control enabling SPM-dominated spectral broadening over large parameter ranges.

In this letter we introduce a post-compression method approaching the few-cycle regime for Yb-laser input pulses using a single gas-filled MPC made of dispersion-engineered dielectric mirrors. We spectrally broaden and post-compress 122 μJ pulses at 1 kHz repetition rate and 150 fs input pulse duration down to 16 fs in a compact (400 mm length) MPC. To overcome limitations set by gas ionization, we optimize the Group Delay Dispersion (GDD) inside the MPC allowing the compression of 250 μJ pulses while preserving the spectral broadening characteristics, resulting in post-compressed pulses of 17 fs.

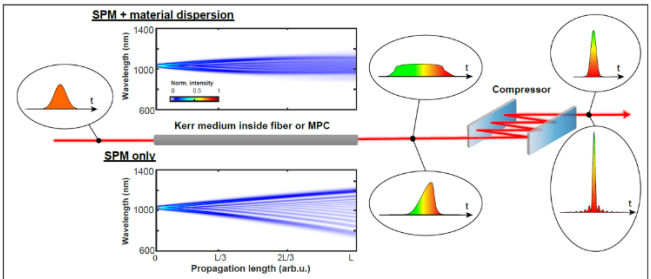

Fig. 1. SPM-based spectral broadening in a normal dispersive medium (top) and dispersion-balanced scheme (bottom): Simulated SPM-based spectral evolution over Kerr medium length $L$ and corresponding temporal pulse intensity profile before and after compression. In both cases, the simulations were performed considering an input pulse duration of 150 fs.

To illustrate the effect of dispersion control on the nonlinear spectral broadening process, we compare two different spectral broadening scenarios considering the same input parameters for both examples, i.e. pulses with 150 fs duration centered at 1030 nm (Fig. 1). In the case of beam propagation through a conventional waveguide, e.g. a gas-filled HCF or MPC made of non-dispersive dielectric mirrors, the broadening process gets saturated rapidly and exhibits strong temporal pulse reshaping. However, in case of an idealized fully dispersion-balanced waveguide, SPM-like spectral broadening is achieved over a larger bandwidth supporting few-cycle pulse durations, while showing only minor temporal pulse reshaping. This enables shorter pulse durations and higher peak intensities albeit exhibiting weak temporal pre- and post-pulses typical for pure SPM [19]. Note that the temporal pulse structure can be cleaner in the first case as discussed in earlier works [20].

Negatively chirped dielectric cell mirrors have been previously used within a solid-state-based MPC for achieving self-compression exploiting the anomalous dispersion regime [21]. However, this approach is limited by the peak intensity that can easily reach the damage threshold of the nonlinear material or the anti-reflection coatings. Here we instead exploit dispersion management for optimizing the spectral broadening process while utilizing high-pressure gases to circumvent the limitation imposed by bulk materials. Via numerical simulations based on a 3D propagation model [22], we investigate different MPC dispersion configurations mimicking the input parameters used for the experiments presented in this work.

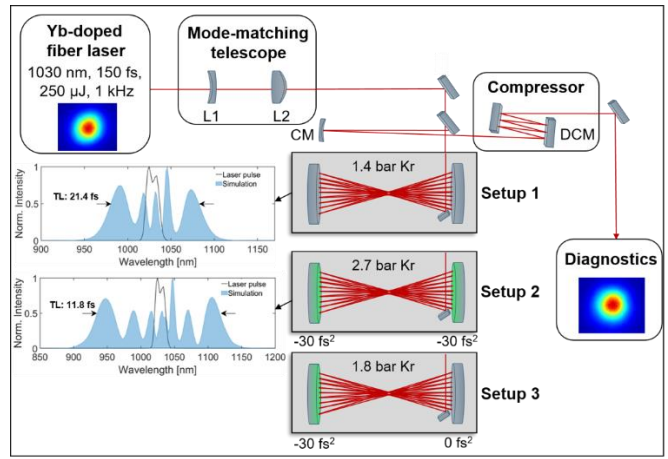

Fig. 2. Schematic of the overall pulse compression setup composed of a mode-matching lens telescope (L1: concave lens and L2: convex lens), a gas-based single-stage MPC, a spherical mirror (CM: concave mirror) for collimation and dispersion-compensating mirrors (DCMs) for post-compression. We distinguish between three different MPC setups: Setup 1 is the conventional configuration with two non-dispersive cell mirrors, Setup 2 incorporates two dispersion-engineered cell mirrors and Setup 3 consists of a combination of one dispersive and one non-dispersive cell mirror. This latter configuration is used to accommodate a higher laser pulse energy. Numerical simulation results based on the 3D propagation model introduced by [22] are displayed for Setup 1 and Setup 2.

As SPM-based spectral broadening is highly dependent on the input pulse shape, the pulse used in the simulations is characterized from the actual laser system to be used in the experiment. The beam profile is assumed to be Gaussian, as no strong spatio-spectral effects are expected in MPCs. In case of a conventional MPC system employing non-dispersive quarter wave stack multi-layer mirrors and considering an input pulse duration of 150 fs at 250 μJ, our simulations predict a transform limited (TL) output pulse duration of 21.4 fs (Fig. 2, setup 1). In contrast, in a dispersion-balanced MPC (dispersion: - 30 fs$^2$ per mirror bounce, pulse energy: 122 μJ), the theoretical prediction indicates a significantly broader spectral bandwidth corresponding to a TL of 11.8 fs (Fig. 2, setup 2). Note that the simulations include the specifications of the mirror coatings, and that the negatively chirped mirrors cover a larger bandwidth than the zero-dispersive ones.

Following our numerical predictions, we experimentally tested the feasibility of our concept. As displayed in Fig. 2, the overall experimental setup consists of a mode-matching lens telescope, a single-stage gas-filled MPC placed inside an overpressure chamber and a chirped-mirror compressor. We employ a commercial Yb-doped fiber laser system (Tangerine, Amplitude) with a nominal pulse duration of 150 fs and a center wavelength of 1030 nm. The pulse energy can be tuned up to 250 µJ at 1 kHz repetition rate with a measured $M^2_{xy}$ of 1.26 x 1.23. The laser pulses are coupled into a compact high-pressure chamber containing two dielectric concave 2" mirrors with a radius of curvature (ROC) of $R$ = -200 mm and arranged in a Herriott-type configuration [23,24] of about $L$ = 400 mm length, i.e. an L/R ratio of roughly 1.98. Two lenses with a focal length of -150 mm and 250 mm, respectively, are used to match the laser beam to the eigenmode of the cell. In- and out-coupling of the beam are realized through one small rectangular dielectric mirror placed in front of one of the cell mirrors. The beam is recollimated by a spherical mirror and sent to a chirped-mirror compressor.

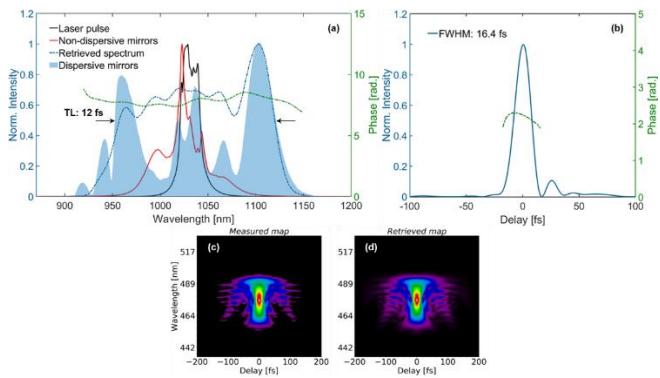

Fig. 3. Output pulse characteristics of the MPC with non-dispersive and dispersive cell mirrors for 122 µJ input pulse energy. (a) Spectral comparison between laser pulse (black), conventional cell configuration (red), MPC configuration with two dispersive cell mirrors with -30 fs$^2$ each (filled blue) and the corresponding FROG-retrieved spectrum (dashed blue) and phase (dashed green). (b) FROG retrieved temporal profile (blue) and phase (dashed green). (c) Measured and (d) retrieved FROG trace.

To achieve nonlinear spectral broadening inside the MPC we distinguish between a conventional and a dispersion-controlled configuration. In the first case, (Fig. 2, Setup 1) standard quarter-wave stack mirrors are used in 1.4 bar krypton and the beam is aligned for 22 round trips through the MPC. The pulse is post-compressed by means of two broadband dispersion-compensating mirrors (DCMs) to compensate for about 1850 fs$^2$ of positive GDD acquired during the nonlinear process and additional 8 mm fused silica glass. In the second case, dispersion-engineered dielectric cell mirrors are employed as MPC mirrors. Two cell mirrors with a GDD of -30 fs$^2$ each are used to compensate linear dispersion during a single pass through the cavity at 2.7 bar krypton and a total of 15 round trips through the MPC (Fig. 2, Setup 2). Post-compression is achieved by employing a broadband DCM pair to compensate for roughly 800 fs$^2$ during propagation inside the MPC.

While for the conventional configuration (Setup 1) the full laser pulse energy (250 uJ) is coupled into the MPC, Setup 2 can only support half of the pulse energy (122 uJ) since the employed dispersive mirrors cause phase overcompensation, and the resulting increased intensity induces ionization. This limit can be easily overcome by reducing the gas pressure to 1.8 bar and by replacing one of the -30 fs$^2$ mirrors by a non-dispersive one (see Setup 3 in Fig. 2), resulting in an approximately dispersion-balanced scenario. Pulse compression is achieved via the same broadband DCM pair which now compensates for a GDD of approximately 600 fs$^2$. The spectra generated from Setups 2 and 3 are acquired with two different spectrometers, Ocean FX and NirQuest by Ocean Insight Inc. in order to detect the full spectral range. A second-harmonic frequency-resolved optical gating (SH-FROG) setup is used to characterize the generated pulses for each configuration. For Setup 1, we reconstruct a pulse duration of 22 fs, in agreement with the transform limit of 21.7 fs (FWHM) of the spectrum represented by the red curve in Fig.3 (a). As explained above, here pulse duration and spectral broadening are not only limited by the bandwidth of the employed dielectric non-dispersive cell mirrors but most importantly by the interplay between linear dispersion of the gas and self-phase modulation (SPM) leading to broadening saturation.

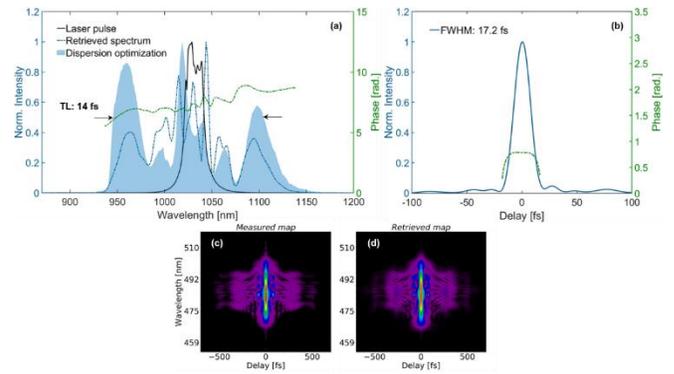

Fig. 4. Output pulse characteristics of the dispersion-optimized MPC for 250 µJ input energy. (a) Spectral comparison between laser pulse (black), MPC configuration with one -30 fs$^2$ and one 0 fs$^2$ cell mirror (filled blue) and its FROG retrieved spectrum (dashed blue) and phase (dashed green). (b) FROG retrieved temporal profile (blue) and phase (dashed green). (c) Measured and (d) retrieved FROG trace.

In contrast, controlling the GDD inside the MPC results in significant additional spectral broadening while maintaining a throughput as high as 98% measured directly at the MPC output. A spectral bandwidth of approximately 130 nm (FWHM) is measured, corresponding to a TL of 12.4 fs at FWHM (Fig. 3 (a), blue filled curve). This is in agreement with the predictions from the numerical simulations. The pulses are characterized and post-compressed down to 16.4 fs with a FROG error of approx. 0.2 %. Both the measured and retrieved FROG trace (Fig.3 (c) and (d)) as well as the retrieved pulse duration (Fig. 3 (b)) show residual third-order dispersion which can be attributed to the DCMs which are not fully optimized for the measured spectral phase. Moreover, the asymmetry of the generated spectrum on the blue side clearly indicates that the cell mirrors coating is imposing a limit to the maximum achievable broadening. As mentioned previously, since only 122 µJ of the laser pulse energy can be compressed when using the MPC in configuration 2, we employed Setup 3 to compress the full laser energy (250 µJ) still maintaining an MPC throughput of more than 90%. Furthermore, the generated spectral bandwidth is preserved and corresponds to a TL of 13.9 fs (Fig. 4(a), blue filled curve). The sharp cut of the spectrum on the blue side can again be attributed to the bandwidth limits of the cell mirror coating. The

FROG-retrieved temporal profile (Fig. 4(b)) shows a compression to 17.2 fs FWHM with a FROG error of approx. 0.3%.

In conclusion, we have demonstrated the potential of direct dispersion control in MPCs to overcome limits of conventional spectral broadening schemes targeting few-cycle pulses. Our approach allows for large compression ratios while maintaining high transmission, excellent beam quality and SPM-dominated spectral broadening even at large spectral bandwidth supporting few-cycle pulse durations. The observed restrictions in pulse duration and spectral bandwidth can be circumvented by further improving the coating design of the cell mirrors.

The here presented approach of precise dispersion engineering has the potential to extend the application range of MPCs even beyond self-phase modulation and towards other nonlinear processes such as soliton [25] and dispersive wave generation [26], four-wave mixing and nonlinear frequency shifting [27].

**Funding.** F.C. acknowledges support from the Deutsche Forschungsgemeinschaft (DFG, German Research Foundation) – SFB-925 – project 170620586 and the Cluster of Excellence Advanced Imaging of Matter (AIM). F.C. and L.S. acknowledge support from the Helmholtz Association via the project HIRS-0018. V.W. acknowledges support from the Partnership for Innovation, Education and Research (PIER) under the PIER seed project PIF-2021-03. A.T. acknowledges support from the Helmholtz association under the Helmholtz Young Investigator Group VH-NG-1613.

**Acknowledgments.** We acknowledge the support of Deutsches Elektronen-Synchrotron (DESY, Hamburg, Germany), Helmholtz-Institute Jena (Germany), members of the Helmholtz Association HGF and Klas Pikull for the great technical support.

**Disclosures.** The authors declare no conflicts of interest.

**Data availability.** Data underlying the results presented in this paper are not publicly available at this time but may be obtained from the authors upon reasonable request.